\documentclass{article}

\usepackage{PRIMEarxiv}

\usepackage[utf8]{inputenc} % allow utf-8 input
\usepackage[T1]{fontenc}    % use 8-bit T1 fonts
\usepackage{hyperref}       % hyperlinks
\usepackage{url}            % simple URL typesetting
\usepackage{booktabs}       % professional-quality tables
\usepackage{amsfonts}       % blackboard math symbols
\usepackage{nicefrac}       % compact symbols for 1/2, etc.
\usepackage{microtype}      % microtypography
\usepackage{lipsum}
\usepackage{fancyhdr}       % header
\usepackage{graphicx}       % graphics
\graphicspath{{media/}}     % organize your images and other figures under media/ folder
\usepackage{siunitx}

\title{MICROFLUIDICS GENERATION OF MILLIMETER-SIZED
MATRIGEL DROPLETS
\thanks{\textit{\underline{Microfluidics Generation of Millimeter-sized Matrigel Droplets}}: 
\textbf{Arnold C, et al.,  DOI:000000/11111.}}
}

\author
{Cory Arnold,$^{1,\#}$; Gabriela Pena Carmona,$^{1,\#}$;
 David A. Quiroz,$^{1,\#}$;\\
Chung X. Thai,$^{1}$; Brenda A.A.B. Ametepe,$^{1}$; I-Hung Khoo,$^{1,2}$; Melinda G. Simon,$^{3}$; Perla Ayala,$^{1,\dag}$; Siavash Ahrar$^{1, \dag}$\\
\\
\normalsize{$^{1}$ Department of Biomedical Engineering, CSU Long Beach, CA, USA}\\
\normalsize{$^{2}$ Department of Electrical Engineering, CSU Long Beach, CA, USA}\\
\normalsize{$^{3}$ Department of Biomedical Engineering, San José State University, CA, USA}\\\\
\normalsize{ \# Authors contributed equally.}\\
\normalsize{ \dag  Corresponding authors.}\\
\normalsize{Last updated May.27.2023}
}

\begin{document}
\maketitle

%The next command sets up an environment for the abstract to your paper.
\newenvironment{sciabstract}{%
\begin{quote}}
{\end{quote}}

\begin{sciabstract}
\begin{center}
\textbf{Abstract}\\
\end{center}

Significant progress has been made to increase access to droplet microfluidics for labs with limited microfluidics expertise or fabrication equipment. In particular, using off-the-shelf systems has been a valuable approach. However, the ability to modify a channel design and, thus, the functional characteristics of the system is of great value. In this work, we describe the development of co-flow microfluidics and their fabrication methods for generating uniform millimeter-sized (0.5 - 2 mm) hydrogel droplets. Two complementary approaches based on desktop CO$_2$ laser cutting were developed to prototype and build durable co-flow droplet microfluidics. After demonstrating the co-flow systems, water-in-oil experiments, and dimensionless number analysis were used to examine the operational characteristics of the system. Specifically, the Capillary number analysis indicated that millimeter-sized droplet generators operated in the desirable geometry-controlled regime despite their length scales being larger than traditional microfluidics systems. Next, the tunable generation of Matrigel droplets was demonstrated. By adjusting the relative flow rates, the droplet size could be tuned. Finally, we demonstrated fibroblast encapsulation and cell viability for up to 7 days as a proof-of-concept experiment. The systems presented are simple and effective tools to generate robust hydrogel droplets and increase the accessibility of this technology to teaching labs or research settings with limited resources or access to microfluidics.
\end{sciabstract}

% Make a new page for the paper
\newpage
% Start introduction
\section{Introduction}

Over the past decade, droplet microfluidics have successfully enabled research across various applications by digitizing assays and reactions \cite{teh2008droplet, chen2022trends}. These applications include single-cell studies \cite{klein2015droplet,brower2020double}, diagnostics \cite{iyer2022advancing}, high throughput screening \cite{agresti2010ultrahigh,price2016discovery}, lab automation\cite{postek2017precise,werner2022phase}, and fluid-driven logic \cite{prakash2007microfluidic}. Commercialized systems that use droplet microfluidics have also emerged (10X Genomics, Bio-Rad ddPCR, and Dolomite) \cite{brower2020double,iyer2022advancing}. More recently, droplet microfluidics have enabled the tunable generation of 3D microenvironments for cell cultures by using hydrogels. This approach has proved useful for the well-controlled generation of tumor spheroids and organoid cultures.\\

To generate 3D cell cultures with microfluidics, various hydrogels, including alginate \cite{tan2007monodisperse,trivedi2010modular,agarwal2017microfluidics}, gelatin methacrylate \cite{shao2022user,zhang2022one}, hyaluronic acid \cite{jeyasountharan2023viscoelastic},and Matrigel \cite{dolega2015controlled,zhang2021microfluidic,wu2020rapid}have been used. Matrigel (a trademark name from Corning) is a mixture of extracellular matrix proteins (including laminin, nidogen, and collagen), growth factors, and heparan sulfate proteoglycans \cite{rojek2022microfluidic}. Matrigel droplets are of great interest for 3D cell culture due to their potential to support physiologically informed 3D environments. In their study, Dolega et al. \cite{dolega2015controlled}, demonstrated effective droplet microfluidics to generate 200-600 $\mu$m Matrigel droplets for epithelial cell cultures in 3D beads for many days. Here beads started as single-cell culture systems and enabled the formation of physiologically relevant features. In another study, Zhang et al. \cite{zhang2021microfluidic}, demonstrated large organoids (500 µm and above) generated by Matrigel droplets through a microfluidic system. The system enabled the realization of large organoids on demand which may be necessary for modeling the behavior and complexity necessary to recapitulate physiological structures and functions of the organoids. Through their approach, the authors successfully maintained the organoids for 1 week.\\

In a comprehensive review, Wang et al. \cite{wang2023recent} summarized the efforts and challenges in applying droplet microfluidics for 3D cell cultures. Specifically, the authors highlighted the need for simpler alternatives to fabricate and control droplet microfluidic systems. Briefly, the methods to develop droplet microfluidics include the soft lithographic technique, laser cutting \cite{werner2022phase}, milling \cite{lashkaripour2018desktop}, and the use of off-the-shelf components \cite{trivedi2010modular,hassani2022developing}. Each method provides unique strengths and potential drawbacks. In this work, simple co-flow microfluidics (a straight channel with two inlets) were developed. Specifically, cleanroom-free fabrication strategies via a desktop laser cutter were used to enable the development of co-flow microfluidics. The co-flow microfluidics demonstrated the tunable generation of mm-sized droplets (0.5-2 millimeter diameter). Water-in-oil experiments were used to characterize the tunability of the droplets. The results from these experiments were combined with a dimensionless analysis to investigate the system’s operating characteristics. Next, the co-flow microfluidics was used to generate Matrigel droplets. Tuning the flow rates enabled the adjustment of the droplet size. Finally, fibroblast cells were cultured using Matrigel droplets for up to 7 days.

\section{Materials and Methods}
Two methods were developed to build millimeter-sized droplet microfluidics (Figure 1). Both methods used a 40 W desktop $CO_{2}$ laser cutter (Glowforge Plus, Glowforge, Seattle, WA). In the first method, channels were directly cut from a plastic sheet and then sealed on both sides with tape. The rapid prototyping approach (about 30 minutes from design to test) were used to investigate the feasibility of designs. Yet, there were issues with leakage from the interface between the inlets and tape for the particular oil and surfactant combination used. Therefore, a second method via laser cutting was developed to create molds for casting durable Polydimethylsiloxane (PDMS) replicas. The second method produced robust channels that could be used for multiple experiments. Here both methods are described.

\subsection{Prototyping Approach} 
The rapid prototyping method was used to investigate the feasibility of a design. Figure 1.A demonstrates the workflow of the approach. Channels were designed using illustration or CAD software. Channel widths ranged from 1-5 mm, and the channel lengths were typically 5-7 cm. Next, using the desktop laser cutter, designs were cut on thermoplastics. The acrylic stock’s thickness set the height of the channel (1/16" height, McMaster-Carr part number 8560K171). Plastic parts were cleaned and carefully sealed via packaging tape (2 to 3.5 mil) on both sides. A self-healing mat and a hand applicator tool (similar to a squeegee used for vinyl cutting to smooth wrinkles) were used to ensure uniform tape application. In the study, up-to three layers of tape were applied to each side of the device to increase mechanical durability. Inlets and outlets were bored inside the top layer of tape using a craft knife or a dispensing needle (27 Gauge or higher). Figure 1. B demonstrates a sample device produced by the prototyping approach.

\subsection{PDMS Microfluidics Fabrication} 

Figure 1.C details the fabrication steps for the PDMS devices. These devices were designed as a single straight channel with two inlets and one outlet. This simple structure formed the co-flow geometry. The designs were prepared with illustration or CAD software. Acrylic sheets (1/16" height) were cut with the 40 W laser cutter. Laser-cut parts were cleaned and then glued to a glass slide to make a mold. Molds were rested overnight to promote adhesion between the glass and plastic. PDMS (Dow Sylgard™ 184 Silicone 0.5 Kg Elastomer Kit) was prepared using the standard protocol, a 10:1 mass ratio of base to curing agent. PDMS replicas were baked and fully cured at 95 $^{\circ}$ C. PDMS replicas were bored using 1 to 2 mm biopsy punches. Finally, PDMS replicas were permanently attached to a glass slide via plasma treatment (Harrick Plasma, Basic Plasma Cleaner, maximum RF power 18W). Figure 1.D demonstrates a sample device produced by this approach.

\subsection{Microfluidics Operation and Cell Culture Viability}

To independently control the volumetric flow rates of the dispersed and continuous phases, two syringe pumps (NE 300 Just Infusion™, New Era Pump Systems inc) were used. The syringe pump for the dispersed phase was placed vertically and above the droplet generator to simplify the tubing connection and prevent cell settlement. The continuous phase was a mixture of oleic acid (food-grade olive oil) and surfactant (Span 80, Sigma Aldrich, 1:100 by volume) for all experiments. Water or Matrigel was used as the dispersed phase. Given the millimeter dimensions of the chambers and droplets, the operation of the device was recorded via a cellular phone camera. The camera was secured in a 10" ring light and tripod stand. Droplet size analysis was conducted with NIH ImageJ. For the cell encapsulation experiments, fibroblast cells (ATCC cells) were premixed with a 5 mg/mL Matrigel solution (Basement Membrane Matrix solution Corning) to create a 2.2 × 105 cells/mL concentration. Care was taken to keep the Matrigel solution and the instruments (chips, tubing, and syringe) cold. Droplets were collected and cultured in a standard 96-well plate with sufficient media. Cell counts were taken at days 3, 5, and 7 of culture to quantify proliferation. To this end, media were removed, and the droplets were washed with 1X PBS before the analysis. Next, nuclei were stained with a DAPI staining solution. Droplets (two for each condition) were visualized with an EVOS fluorescence microscope, and a previously described automated protocol was used for cell count \cite{Labno_cell_count}.

\section{Results}
\subsection{Rapid prototyped chambers}
First, prototyped channels were used as proof-of-concept experiments to investigate the feasibility of the proposed approach. Specifically, straight channels (width 1.5 mm, length 5 mm, and height 3.18 mm) were used as a co-flow droplet generator. Devices were primed with oil before each experiment. Oil with surfactant was used as the continuous phase. The continuous phase was delivered with a tube (1/16" inner diameter, McMaster-Carr, part number 5454K74). Water with trace food dye was used as the dispersed phase, delivered via polyethylene tubing (PE20, inner diameter 0.38 mm and outer diameter 1.09 mm). The tubes were placed directly inside the channels. Two tubes were separated from each other by 1.5 cm along the length of the devices.

To examine the droplet properties as a function of flow rates, first, the volumetric flow rate of the continuous phase was kept constant at Q.oil= 3.5 mL/hr. Four volumetric flow rates for the dispersed phase were examined (1, 1.5, 2, and 2.5 mL/hr. Figure 2 demonstrates the relationship between the droplet diameter and the flow rate ratios. \\
The flow rate ratio was defined as:

\begin{center}
\emph{Equation-1:}  $\phi = Q.ratio = \frac{Q.oil}{Q.water}$
\end{center}

Increasing the flow ratio decreased the droplet diameter. Additionally, the droplet uniformity improved as the flow ratio increased. Specifically, the coefficient of variance (COV) for $\phi = 1.4$ (the lowest value) was 14\%. In contrast, the COVs for all other conditions were less than 7\%. These experiments successfully demonstrated the feasibility of generating droplets with straight co-flow channel geometry. The essential advantage of these experiments was the rapid turnover time from a design to a testable prototype. On average, this task could be accomplished in under 30 minutes. However, there were issues related to oil leaking from the inlet due to the lack of mechanical durability of the tape. Therefore, it was necessary to translate the tape-sealed designs to PDMS cast channels.

\subsection{PDMS Cast Channels and Water-in-Oil-Droplets}
First, water-in-oil experiments was investigated to characterize the operation of the droplet generators. In these experiments, channels were 1.6 mm tall and 4 mm wide. For all conditions, the volumetric flow rate of the oil was kept constant at 5 mL/hr. The volumetric flow rate of water (dispersed phase) was 0.5, 1, 1.5, 2, and 3 mL/hr. All conditions produced droplets.
Figure-3 demonstrates the relationship between the droplet diameter and the flow rate ratios for the durable PDMS channels. The flow rate ratio was defined previously via equation-1. Droplet diameter (here defined as the length of the droplet) could be tuned by adjusting the ratio of volumetric flow rates. Increasing the flow ratio led to more uniform droplets. The coefficient of variance for all conditions was less than 10\%, which suggested good uniformity.\\
Next, the results from these experiments were used as part of a dimensionless number analysis. Specifically, the Capillary number (Ca\#) and Weber number (We\#) were considered. The Capillary number compares the ratio of the viscous shear forces provided by the continuous phase to the interfacial tension forces, which hold together the dispersed phase fluid as it emerges into the channel. The Weber number compares the ratio of inertial forces to interfacial tension forces.\\

The Capillary number ($Ca\#$) was calculated by:

\begin{center}
\emph{Equation-2:}  $Ca\# = \frac{\mu U}{\gamma}$
\end{center}

where $\mu$ is the dynamic viscosity of a phase,\\
U is the average velocity of a phase, and\\
$\gamma$ is the interfacial tension between the dispersed and continuous phases.\\

The Weber number ($We\#$) was calculated by:

\begin{center}
\emph{Equation-3:}  $We\# = \frac{\rho U^2 a_0}{\gamma}$
\end{center}

where $\rho$ is the density of a phase,\\
U is the average velocity of the dispersed phase,\\
$ a_0 $ is the characteristic length scale of the system taken as the channel width, and\\
$\gamma$ is the the interfacial tension between the dispersed and continuous phases.\\

A Jupyter notebook was developed to simplify the analysis. The notebook and other visualization tools are available from the project Open Science Framework (OSF) titled Flow Milli. The page can be retrieved from /url{https://osf.io/fcr8w/}. Results from the analysis are provided in Table - 1. Briefly, the Capillary number for the continuous oil phase was $3.96 \times 10^ {- 4}$. The Capillary number for the dispersed phase ranged from  $1.48 \times 10^{-6}$ to  $8.9 \times 10^ {- 6}$. Comparing the Capillary numbers with previously described frameworks suggested that droplets were generated in the desirable geometry-controlled regime \cite{anna2016droplets}. Moreover, the Weber number for the dispersed phase ranged from $1.23 \times 10^ {- 7}$ to $4.93 \times 10^ {- 6}$, suggesting that interfacial tension (similar to micrometer-sized chambers) dominated the system. Together the water-in-oil experiments provided a helpful framework for the subsequent Matrigel droplets investigations.

\subsection{Matrigel-in-Oil-Droplets}
Next, tunable generation of hydrogel droplets was investigated. Matrigel was selected due to its prolific use in 3D cell culture experiments, including organoids and spheroids. To conserve Matrigel, channels with a 1.5 mm width were used. In these experiments, the volumetric flow rate of the continuous phase was kept at 3.5 mL/hr. The volumetric flow rates for the dispersed phase included 0.25, 0.5, 0.75, 1, and 1.25 mL/hr. Figure-4 summarizes the relationship between the Matrigel droplet size and the flow rate. Similar to the water droplets, increasing the flow ratio led to smaller and more uniform droplets. However, Matrigel droplets were less tunable than water droplets. Both the hydrogel’s rheological properties and the chamber’s smaller width (1.5 mm vs. 4 mm) may have contributed to this outcome.

\subsection{Cell-culture Feasibility}
Next, Matrigel droplets were used for cell culture as a proof of concept experiment. In these experiments, the Matrigel flow rate was fixed at 0.5 mL/hr. The flow rate was selected since it produced the smallest coefficient of variance (COV) and provided the highest possibility for maintaining droplet integrity. A 1\% Matrigel solution was prepared as described in supplemental materials. Cells were mixed with the Matrigel solution. Droplets were collected in a 96-well plate (containing media). Cell proliferation measurements were conducted on days 3, 5, and 7. For each time point, 4 independent wells, each containing 2 droplets, were used (Figure 5). While preliminary, these experiments supported the feasibility of cell culture for up to 7 days within the Matrigel droplets. Unfortunately, there was some loss of the droplet shape during the transfer of the droplets to the 96-wells. Therefore, additional optimizations are required to ensure the integrity of the droplets in future studies. Despite the poor stability of the droplets, the 3D hydrogel environment maintained cell viability.

\section{Discussion}
Droplet microfluidics has emerged as a powerful platform for generating 3D cell cultures \cite{wang2023recent}. Significant progress has been made to provide broader access to droplet microfluidics for labs without microfluidics expertise. In particular, using off-the-shelf systems has been valuable \cite{trivedi2010modular}. However, the ability to modify a channel design and, thus, the functional characteristics of the system is of great value. Approaches (3D printing, CNC milling, and laser-cutting) complementary to conventional soft-lithography have enabled broader access to building microfluidics. This work described two complementary approaches based on desktop laser cutting to prototype and develop durable co-flow droplet microfluidics. Desktop laser cutters are available to most labs through the existing maker space infrastructures in a university context. In the prototyping approach, the feasibility of designs was rapidly assessed. In the second approach, laser-cut plastic parts were used as a mold to build durable microfluidic chambers. Together both methods provide an approach to iterate quickly through microfluidic designs based on a project’s needs. Moreover, millimeter size dimensions, instead of conventional micrometer chambers, were developed to enable tissue engineering applications such as fabrication of larger 3D microenvironments with sufficient volumes, necessary for development of more complex multicellular structures and characteristics of tumors, such as formation of a necrotic core. The operational characteristics of the systems were examined using water-in-oil experiments and dimensionless analysis. The Capillary number indicated that co-flow systems behaved in the desirable geometry-controlled regime. However, it is essential to highlight that the experiments and analysis used an idealized dispersed phase (water) as opposed to Matrigel.

The co-flow microfluidics provides a platform to systematically examine droplet generation with various hydrogels (e.g., polyethylene glycol, alginate, and hyaluronic acid). Moreover, two additional parameters can be considered in the system performance. First, the hydrophobic properties of the chambers could be improved. The use of tape or adhesive to generate hydrophobic droplet generators has been previously demonstrated \cite{thompson2013adhesive}. Similarly, rapidly prototyped channels demonstrated ideal hydrophobic properties suitable for aqueous-in-oil droplets. No hydrophobic treatments were used for the experiments with the durable chambers. In the future, we plan to investigate using more hydrophobic substrates, for example, plastic or PDMS parts instead of a glass slide, to create the droplet generators. Second, the tube’s shape or geometry delivering the dispersed phase could be improved. In this investigation, tubes were directly placed inside the channels. While simple, this approach limited the ability to systematically investigate the role of the nozzle shape in the generation of droplets. In the future, off-the-shelf connectors or 3D-printed parts could enable the modification of this feature in the system. Together, these modifications could provide better control towards the uniform generation of hydrogel droplets. In particular, a more robust tubing-chip connection could increase the throughput (frequency of droplet production) by enabling maintenance of the back-pressures needed to sustain leak-free connections at higher flow rates.

\section{Conclusion}
In this study, millimeter-sized droplets were generated with co-flow droplet microfluidics. To this end, a simple fabrication approach based on laser cutting was demonstrated. Water-in-oil droplet experiments enabled characterization and subsequent analysis of the system. Specifically, the Capillary number analysis indicated that the devices operated in the geometry-controlled regime. Next, the tunable generation of Matrigel droplets was investigated. The Matrigel droplets verified successful fibroblast cell encapsulation and cell proliferation for up to seven days. While tunable, Matrigel droplets were less uniform than the water droplets. Additionally, maintaining the droplet mechanical integrity inside the 96 well-plate presented challenges. In future experiments, adjustments can be explored, including maintaining the cells inside a tubing or microfluidic traps \cite{trivedi2010modular,zhang2021microfluidic}. Alternatively, droplets can be maintained inside other hydrogel environments instead of liquid media. Tunable generation of millimeter-sized hydrogel droplets could provide a versatile approach for scaling biomanufacturing, 3D cell-cultures, and microphysiological applications.

\newpage
\section* {Acknowledgments}  
Authors express gratitude to Dr. Philip Thomas, members of Ahrar-Lab (in particular Michael S. Richardson) and Prof. Elliot Hui for helpful discussions during the project. \\
This work was supported in part by CSULB startup funds and a CSUPERB new investigator grant to SA. Additionally, this work was supported by National Institutes of General Medical Sciences to PA.
GPC and DAQ were supported by the CSULB’s NIH BUILD Scholars program.\\

\section* {Conflicts of Interest} 
The authors declare no conflict of interest.

\section* {Additional Resources}
Additional resources (code, bill of materials, etc)
are available from the project OSF public repository titled: \emph{Flow-Milli}, which can be retrieved from:\url{https://osf.io/fcr8w/}

\section*{Corresponding Authors Addresses}
Siavash Ahrar (Ph.D.)\\
Mail: Department of Biomedical Engineering, CSU Long Beach, 
1250 Bellflower Blvd. \\VEC-404.A, Long Beach, CA 90840, 
E-mail: Siavash.ahrar@csulb.edu\\

Perla Ayala (Ph.D.)\\
Mail: Department of Biomedical Engineering, CSU Long Beach, 1250 Bellflower Blvd.\\
ET-108, Long Beach, CA 90840 E-mail: Perla.ayala@csulb.edu\\

\textbf {List of Figures and Tables}:
\begin{itemize}  
  \item Figure-1: Fabrication protocols for millimeter-sized droplet microfluidics.
  \item Figure-2: Rapidly prototyped water-in-oil millimeter sized droplets.
  \item Figure-3: PDMS replicated water-in-oil millimeter sized droplets.
  \item Figure-4: Matrigel-in-oil millimeter sized droplets.
  \item Figure-5: Cell culture feasibility experiments.
  \item Table-1: Dimensionless number analysis for water-in-oil experiments via PDMS replicated devices.
\end{itemize}

\newpage
\section{Supplemental Materials}
\subsection{Matrigel Preparation}

Matrigel Basement Membrane matrix (Corning) thawed overnight or 3 hours before usage. The syringe and pipette tips were placed inside a lab freezer to ensure the solution remained cold during the experiment. Matrigel was diluted to a final 5.0 mg/mL concentration with media. Matrigel was diluted in media using the following calculations:\\

\begin{center}
$C1 \times V1 = C2 \times V2 $
\end{center}

C1: Initial concentration of Matrigel = 8.9 mg/mL\\
C2: Final concentration of Matrigel = 5.0 mg/mL\\
V2: Total volume=810 $\mu$L.

Therefore, the volume of Matrigel needed:
V1 = 455 $\mu$L.\\

810 $\mu$L Total - 455 $\mu$L Matrigel = 354.9 $\mu$L Media or Cell Solution.

To prepare the necessary volume for Matrigel concentration dilution, fibroblasts (ATCC) were passaged using a standard protocol. The cells were seeded at a density of 500,000 cells/mL.\\

Volume of cell solution:

\begin{center}
$\frac{500,000. cells}{9,905,000. cells} \times 7 mL = 0.35 mL = 353.35 \mu L $
\end{center}

The 8.9 mg/mL initial Matrigel stock was diluted to 5 mg/mL using the fibroblasts cell solution.

\newpage
\begin{flushleft}
\bibliographystyle{unsrt}
\bibliography{refs} % Entries are in the "refs.bib" file
\end{flushleft}

\newpage

\begin{figure}[b]
\includegraphics[width=\textwidth]{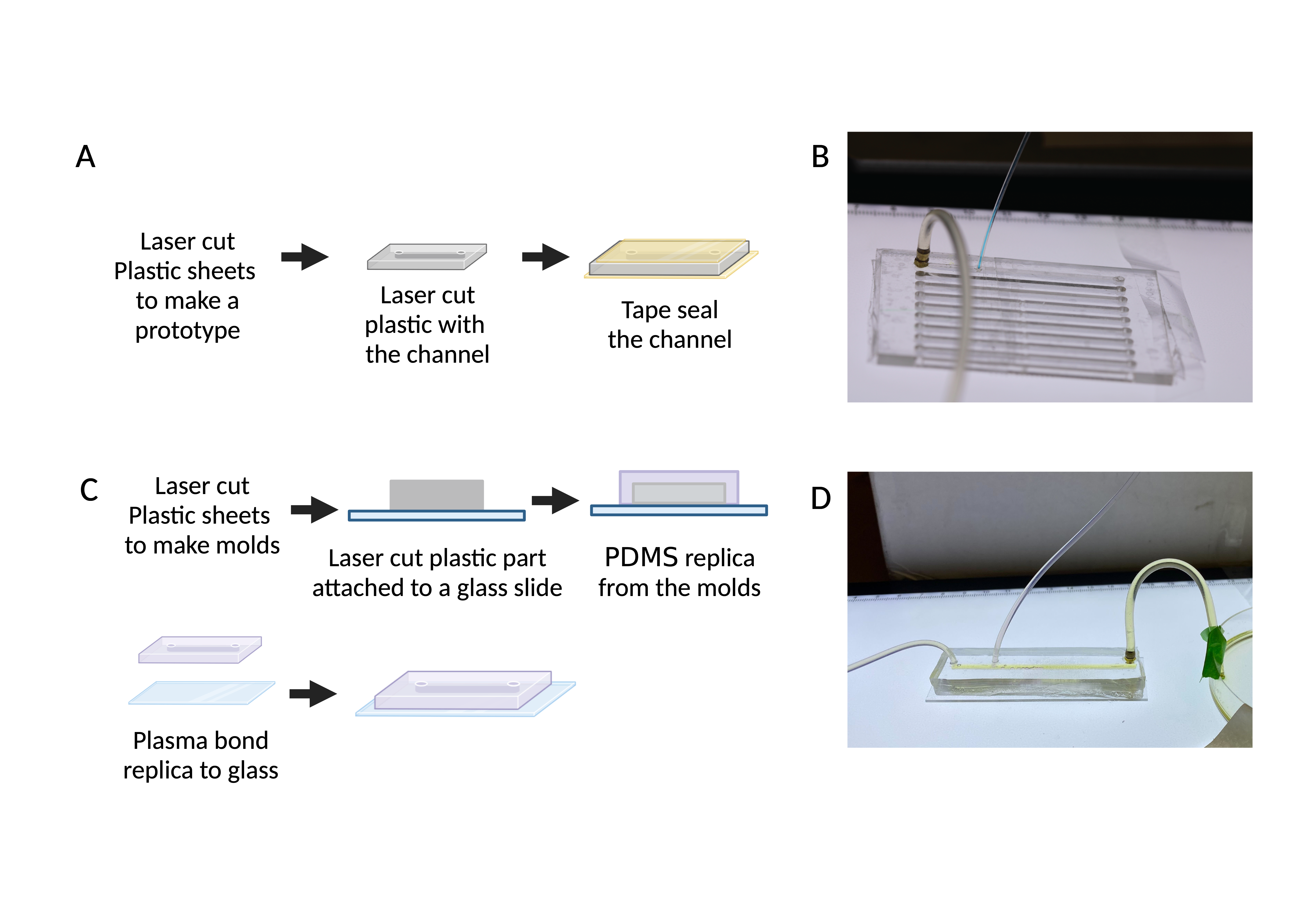}
\centering
\caption{\textbf {Fabrication protocols for millimeter-sized droplet microfluidics}. (A) Rapid prototyping approach. Designs were first prototyped with laser cut plastic channels which were then tape sealed. (B) An example of devices generated with the rapid prototype approach. (C) Durable PDMS device manufacturer. Laser cut plastic parts were used to make a mold. PDMS replica were cast from the mold and plasma bonded to glass to form the channel. (D) An example of finalized microfluidic device that were PDMS cast from laser-cut molds. Figure in part produced by BioRender.}
\end{figure}

\begin{figure}[b]
\includegraphics[width=\textwidth]{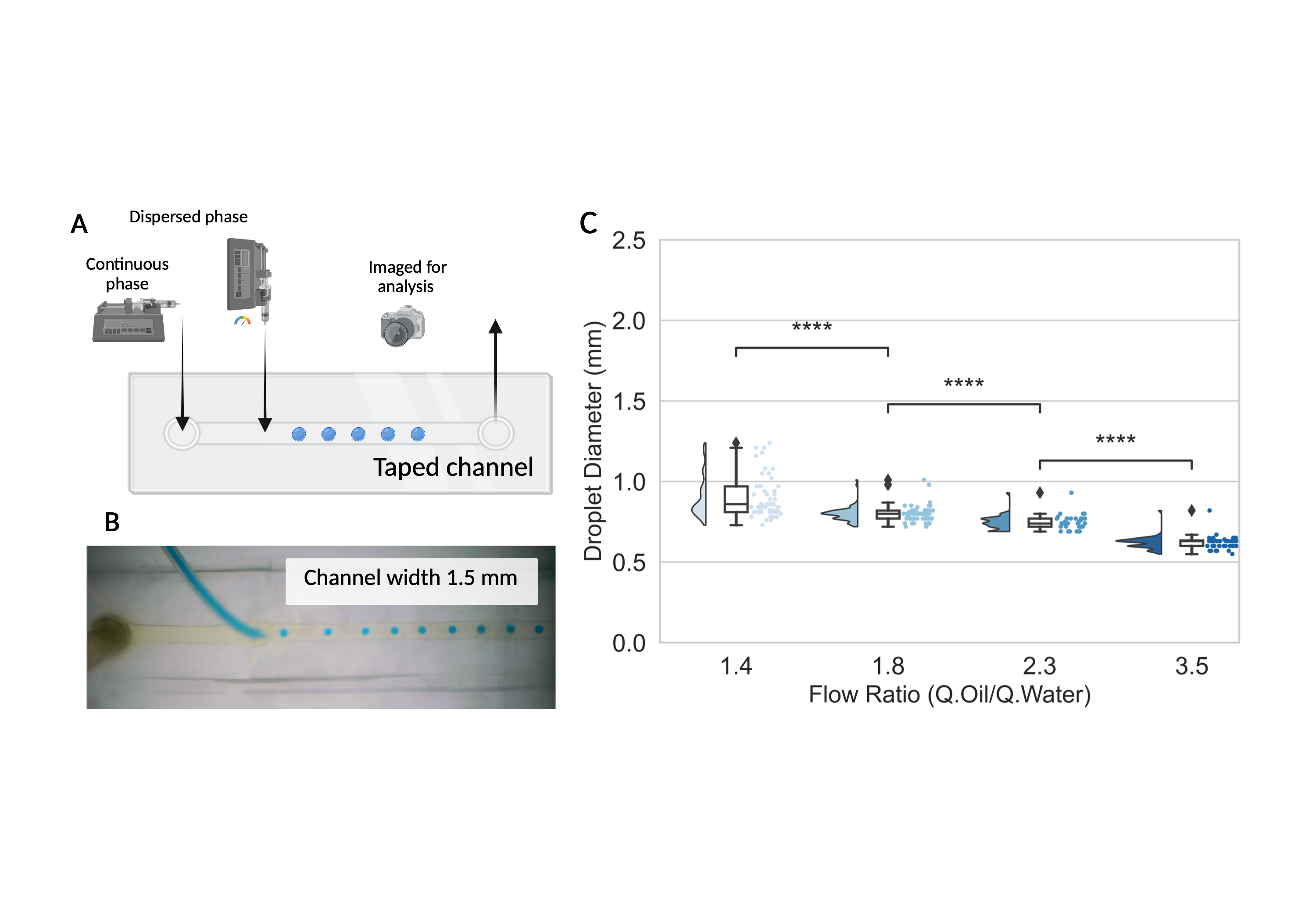}
\centering
\caption{\textbf {Rapidly prototyped water-in-oil millimeter sized droplets.}  (A) Laser cut plastic channels were tape sealed to form a prototype. (B) Characterization of the rapid-prototyped droplet generators with water-in-oil experiments. (C) Increasing the flow ratio decreased the droplet size and improved uniformity. The volumetric flow rate for the continuous phase (oil) was kept constant at 3.5 mL/hr for all conditions. Figure in part produced by BioRender.}
\end{figure}

\begin{figure}[b]
\includegraphics[width=\textwidth]{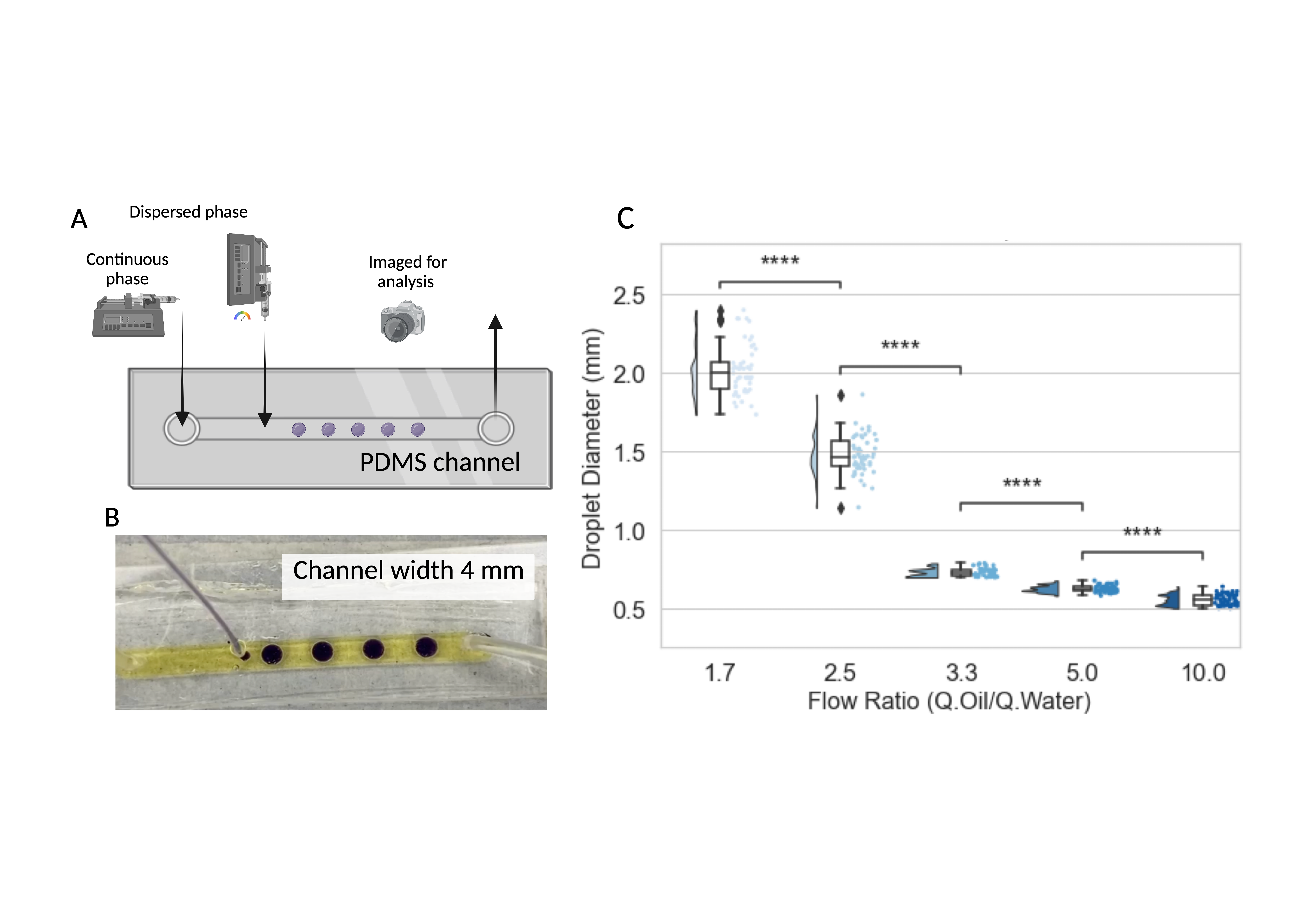}
\centering
\caption{\textbf {PDMS replicated water-in-oil millimeter sized droplets}. (A) Co-flow microfluidics -a straight channel with two inlets and one outlet- produced tunable droplets. (B) Image of the device operation. The channel width was 4 mm. Continuous flow rate was fixed at 5 mL/hr. (C) Droplet size (0.5 - 2 mm) could be tuned by adjusting the flow ratios. Increasing the flow ratio reduced the droplet size and increased uniformity. The coefficient of variance (COV) for all conditions was less than 10\%.  Figure in part produced by BioRender.}
\end{figure}

\begin{figure}[b]
\includegraphics[width=\textwidth]{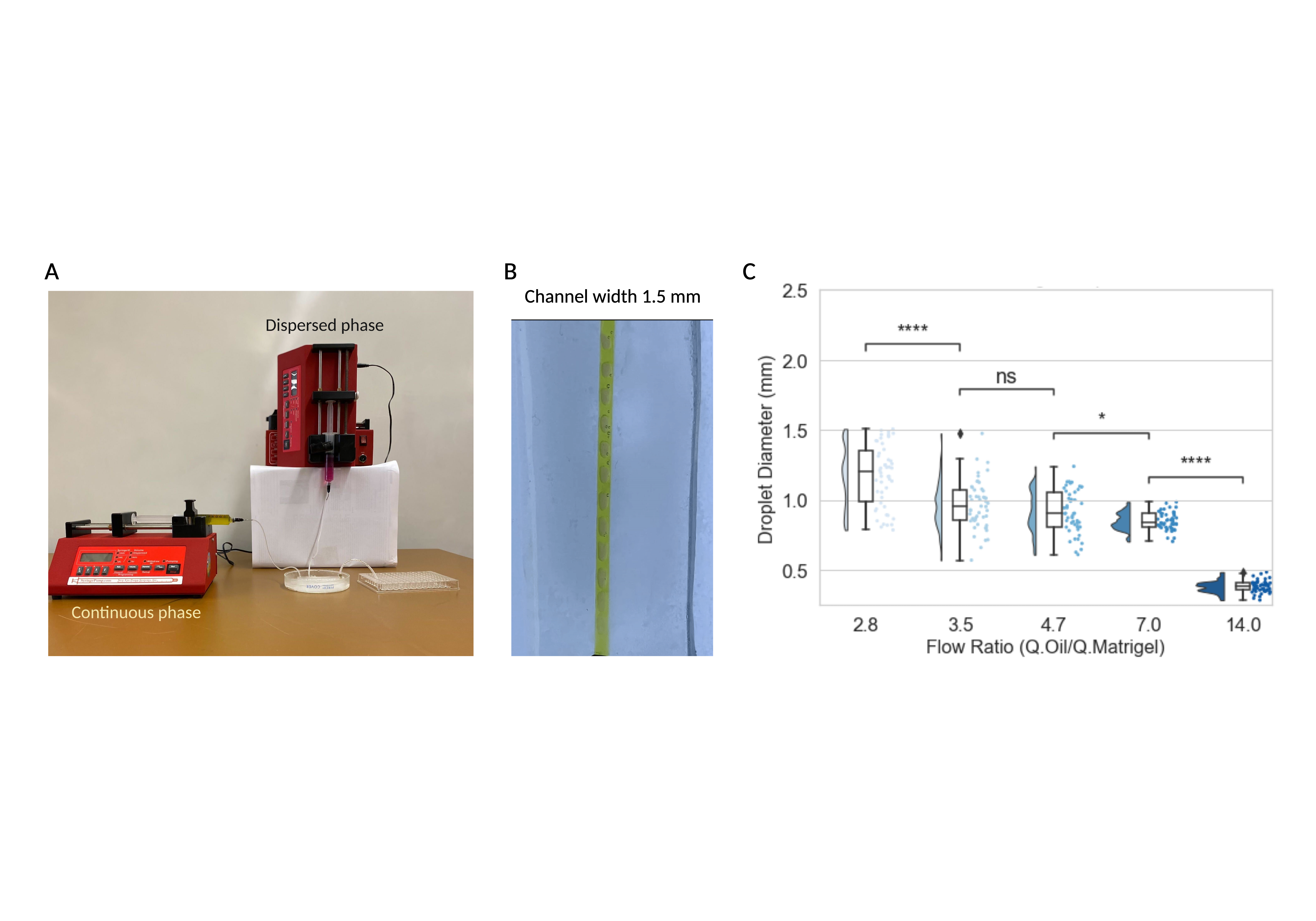}
\centering
\caption{\textbf {Matrigel-in-oil millimeter sized droplets.} (A) Set up for the Matrigel experiments. The chip was placed on an ice puck to keep the chamber cold. (B) Co-flow microfluidics channel width 1.5 mm. (C) Increasing the flow ratio improved the droplet uniformity. Matrigel droplets were less tunable than the water droplets.}
\end{figure}

\begin{figure}[b]
\includegraphics[width=\textwidth]{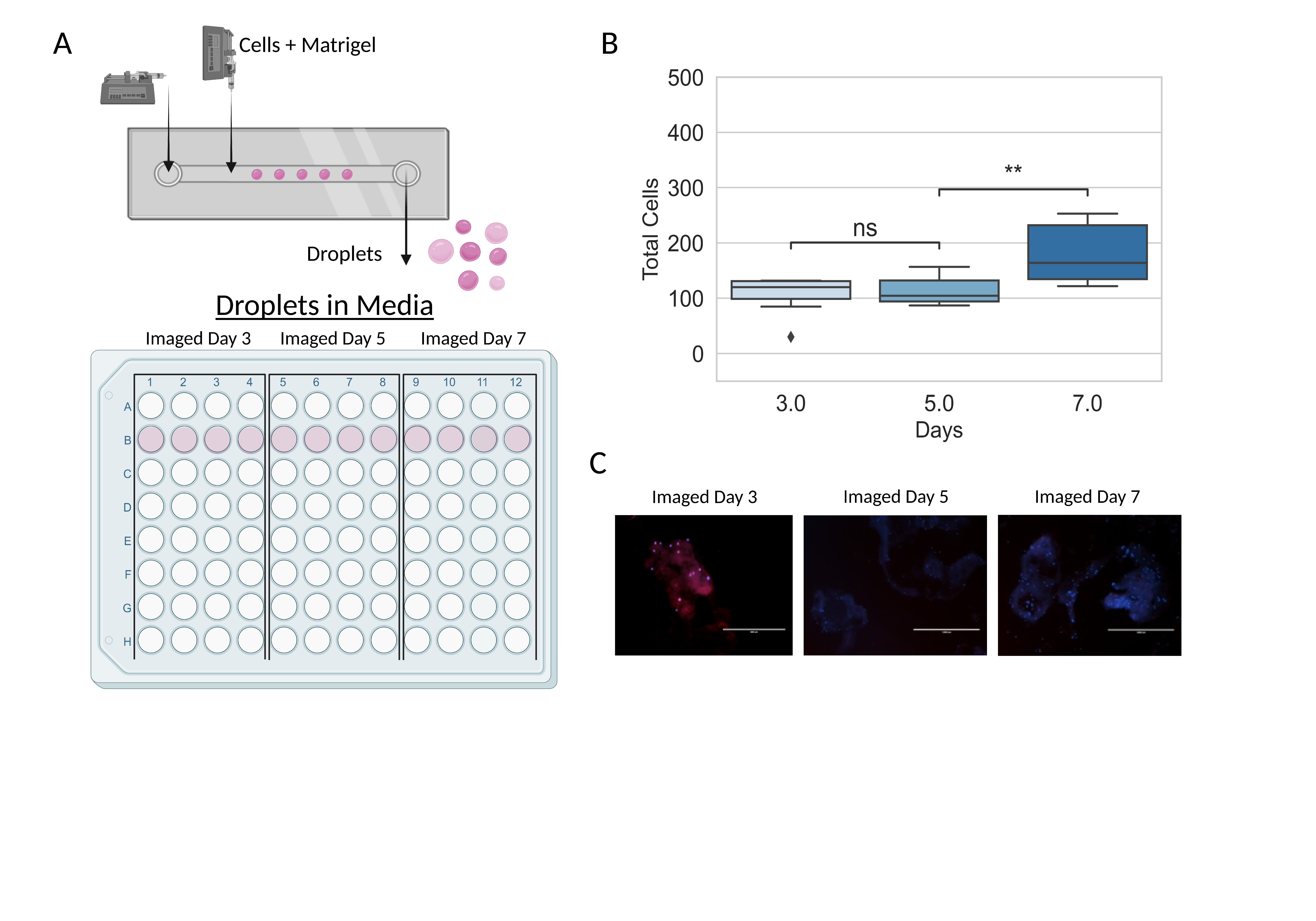}
\centering
\caption{\textbf {Cell culture feasibility experiments.} (A) Fibroblast cells were pre-mixed with 5mg/mL Matrigel solution and used for droplet generation. Matrigel flow rate was fixed at 0.5 mL/hr. Droplets were collected and cultured with media. For each condition, two Matrigel droplets were placed in a well. Four wells for each condition were used, bringing the total to eight droplets. (B) Cell proliferation over 7 days. (C) Representative DAPI+ fibroblasts encapsulated in Matrigel droplets for each respective day. Figure in part produced by BioRender.}
\end{figure}

\begin{figure}[b]
\includegraphics[width=\textwidth]{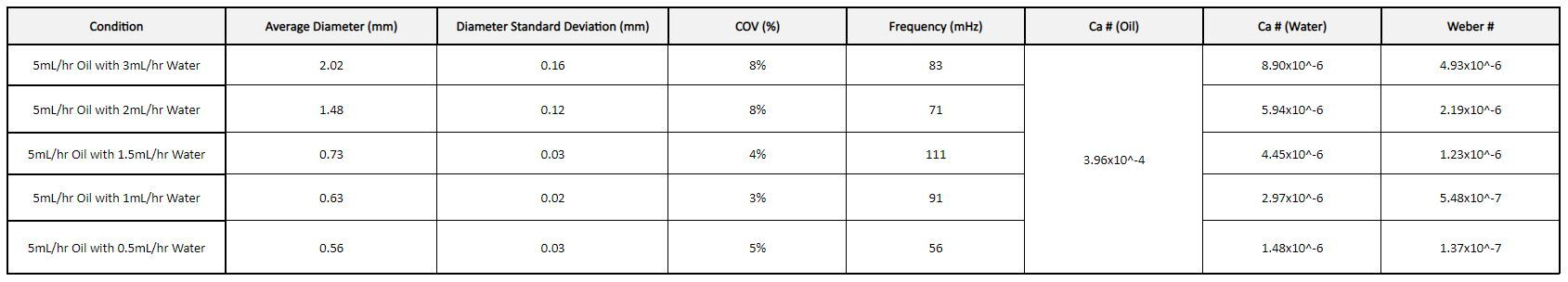}
\centering
\caption{\textbf {Table-1: Dimensionless number analysis for water-in-oil experiments via PDMS replicated devices.}}
\end{figure}

\end{document}